\documentclass[twocolumn]{revtex4}

\newcommand{\be}{\begin{eqnarray}}
\newcommand{\ee}{\end{eqnarray}}
\newcommand{\ket}[1]{\ensuremath{\left| {#1} \right>}}
\newcommand{\bra}[1]{\ensuremath{\left< {#1} \right|}}

\newcommand{\create}{\ensuremath{{\,\hat{a}^{\dagger}}}}
\newcommand{\destroy}{\ensuremath{\,\hat{a}}}

\usepackage{latexsym}
\usepackage{graphicx}
\usepackage{amssymb}
\usepackage{multirow}
\usepackage{textcomp}

\begin{document}
\title{Spin-motion entanglement and state diagnosis with squeezed oscillator wavepackets}

\author{Hsiang-Yu Lo$^\ast$, Daniel Kienzler, Ludwig de Clercq, Matteo Marinelli, Vlad Negnevitsky, Ben C. Keitch, \\ Jonathan P. Home }

\email{hylo@phys.ethz.ch}
\email{jhome@phys.ethz.ch}

\affiliation{Institute for Quantum Electronics, ETH Z\"urich, Otto-Stern-Weg 1, 8093 Z\"urich, Switzerland}

\maketitle

\textbf{Mesoscopic superpositions of distinguishable coherent states provide an analog to the Schr\"odinger's cat thought experiment \cite{13Wineland, 13Haroche}. For mechanical oscillators these have primarily been realised using coherent wavepackets, for which the distinguishability arises due to the spatial separation of the superposed states \cite{96Monroe, 07McDonnell, 04Haljan}. Here, we demonstrate superpositions composed of squeezed wavepackets, which we generate by applying an internal-state dependent force to a single trapped ion initialized in a squeezed vacuum state with 9 dB reduction in the quadrature variance. This allows us to characterise the initial squeezed wavepacket by monitoring the onset of spin-motion entanglement, and to verify the evolution of the number states of the oscillator as a function of the duration of the force. In both cases, we observe clear differences between displacements aligned with the squeezed and anti-squeezed axes. We observe coherent revivals when inverting the state-dependent force after separating the wavepackets by more than 19 times the ground state root-mean-square extent, which corresponds to 56 times the r.m.s. extent of the squeezed wavepacket along the displacement direction. Aside from their fundamental nature, these states may be useful for quantum metrology \cite{02Munro} or quantum information processing with continuous variables \cite{12Weedbrook, 00Gottesman2, 02Bartlett}.}

The creation and study of nonclassical states of spin systems coupled to a harmonic oscillator has provided fundamental insights into the nature of decoherence and the quantum-classical transition. These states and their control form the basis of experimental developments in quantum information processing and quantum metrology \cite{95Monroe2, 13Wineland, 13Haroche}. Two of the most commonly considered states of the oscillator are squeezed states and superpositions of coherent states of opposite phase, which are commonly referred to as ``Schr{\"o}dinger's cat'' (SC) states. Squeezed states involve reduction of the fluctuations in one quadrature of the oscillator below the ground state uncertainty, which has been used to increase sensitivity in interferometers \cite{11Ligo,13Aasi}. SC states provide a complementary sensitivity to environmental influences by separating the two parts of the state by a large distance in phase space. These states have been created in microwave and optical cavities \cite{13Vlastakis, 13Haroche}, where they are typically not entangled with another system, and also with trapped ions \cite{96Monroe, 07McDonnell, 04Haljan, 13Wineland}, where all experiments performed have involved entanglement between the oscillator state and the internal electronic states of the ion. SC states have recently been used as sensitive detectors for photon scattering recoil events at the single photon level \cite{13Hempel}.

In this Letter, we use State-Dependent Forces (SDFs) to create superpositions of distinct squeezed oscillator wavepackets which are entangled with a pseudo-spin encoded in the electronic states of a single trapped ion. We will refer to these states as Squeezed Wavepacket Entangled States (SWES) in the rest of the paper. By monitoring the spin evolution as the entanglement with the oscillator increases \cite{05Zeng, 10Gerritsma, 12Casanova}, we are able to directly observe the squeezed nature of the initial state. We obtain a complementary measurement of the initial state by extracting the number state probability distribution of the displaced-squeezed states which make up the superposition. In both measurements we observe clear differences depending on the force direction. We show that the SWES are coherent by reversing the effect of the SDF, resulting in recombination of the squeezed wavepackets, which we measure through the revival of the spin coherence.

\begin{figure}[b!]
\includegraphics[width = .95\columnwidth]{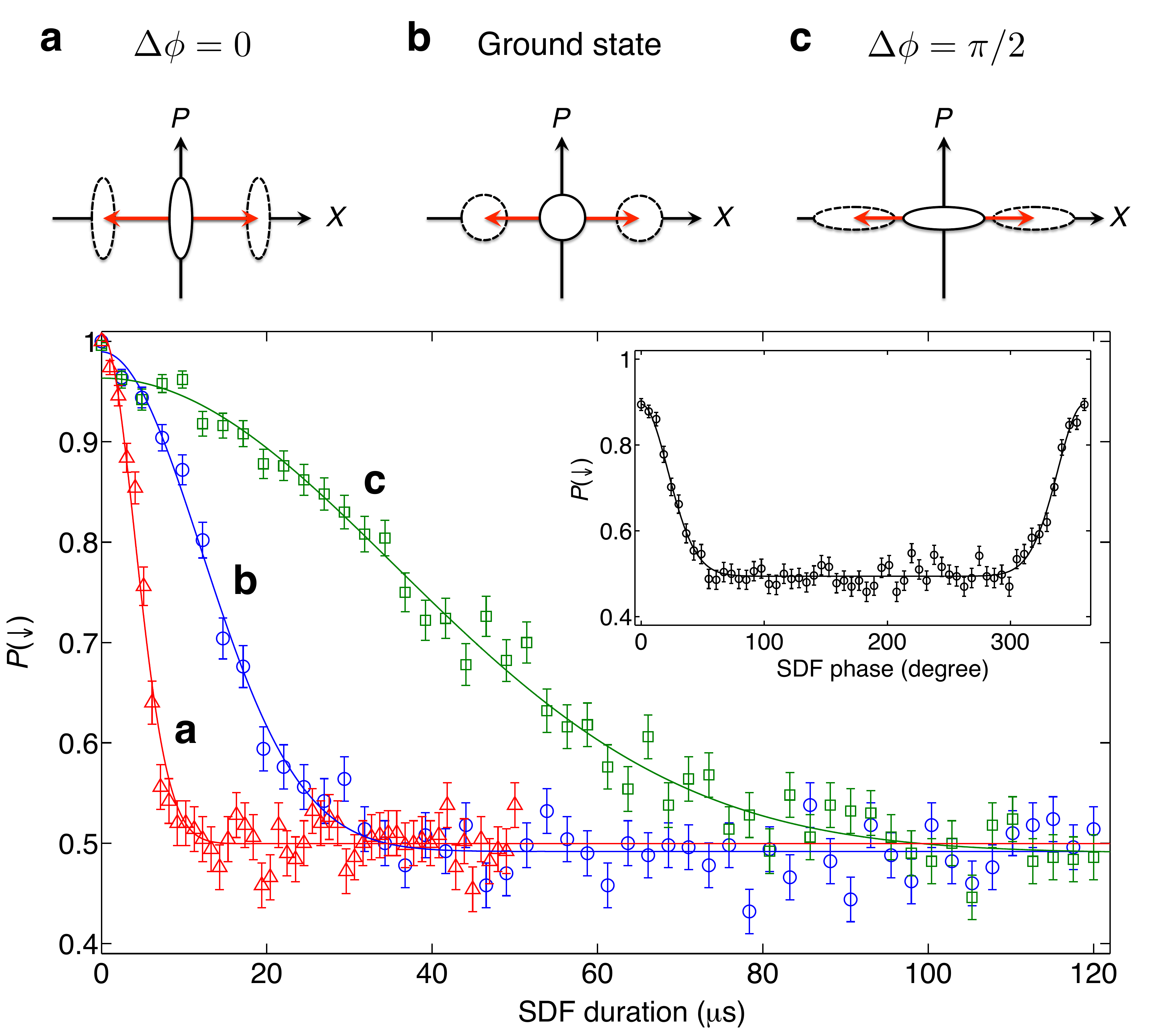}
\caption{\textbf{Spin-population evolution due to spin-motion entanglement:} Projective measurement of the spin in the $\hat{\sigma}_z$ basis as a function of SDF duration. \textbf{a}, Forces parallel to the squeezed quadrature (red triangles). \textbf{b}, An ion initially prepared in the motional ground state (blue circles). \textbf{c}, Forces parallel to the anti-squeezed quadrature (green squares). The inset shows a scan of the phase of the SDF for an initial squeezed state with the force duration fixed to 20~$\mu$s. Each data point is the result of $>300$ repetitions of the experimental sequence. The given error bars indicate one standard error of the mean, and are generated under the assumption that the dominant source of fluctuations is quantum projection noise.}
\label{fig:stateoverlap}
\end{figure}

The squeezed vacuum state $\ket{\xi}$ is defined by the action of the squeezing operator $\hat{S}(\xi) = e^{(\xi^*\destroy^2 - \xi\create^2)/2}$ on the motional ground state $\ket{0}$, where $\xi = r e^{i \phi_s}$ with $r$ and $\phi_s$ real parameters which define the magnitude and the direction of the squeezing in phase space. To prepare squeezed states of motion in which the variance of the squeezed quadrature is reduced by around 9~dB relative to the ground state wavepacket we utilize reservoir engineering, in which a bichromatic light field is used to couple the ion's motion to the spin states of the ion which undergo continuous optical pumping. This dissipatively pumps the motional state of the ion into the desired squeezed state, which is the dark state of the dynamics. More details regarding the reservoir engineering can be found in \cite{14Kienzler}. This approach provides a robust basis for all experiments described below, typically requiring no re-calibration over several hours of taking data. In  the ideal case, the optical pumping used in the reservoir engineering results in the ion being pumped to $\ket{\downarrow}$. To create a SWES, we apply a SDF to this squeezed vacuum state by simultaneously driving the red $\ket{\downarrow}\ket{n}\leftrightarrow \ket{\uparrow}\ket{n - 1}$ and blue $\ket{\downarrow}\ket{n}\leftrightarrow \ket{\uparrow}\ket{n + 1}$ motional sidebands of the spin flip transition \cite{96Monroe}. The resulting interaction Hamiltonian can be written in the Lamb-Dicke approximation as
\be
\hat{H}_D = \hbar \frac{\Omega}{2} \hat{\sigma}_x \left( \create e^{-i \phi_D/2} + \destroy e^{i \phi_D/2}\right)  \ ,
\ee
where $\Omega$ is the strength of the SDF, $\phi_D$ is the relative phase of the two light fields, and $\hat{\sigma}_x \equiv \ket{+}\bra{+} - \ket{-}\bra{-}$ with $\ket{\pm} = (\ket{\uparrow} \pm \ket{\downarrow})/\sqrt{2}$.
For an ion prepared in $\ket{+}$, this Hamiltonian results in displacement of the motional state in phase space by an amount $\alpha(\tau) = -i \Omega e^{-i\phi_D/2} \tau/2$ which is given in units of the r.m.s. extent of the harmonic oscillator ground state. An ion prepared in $\ket{-}$ will be displaced by the same amount in the opposite direction. In the following equations, we use $\alpha$ in place of $\alpha(\tau)$ for simplicity. Starting from the state $\ket{\downarrow} \ket{\xi}$, application of the SDF ideally results in the SWES
\be
\ket{\psi(\alpha)} = \frac{1}{\sqrt{2}}\left(\ket{+} \ket{\alpha, \xi} - \ket{-} \ket{-\alpha, \xi} \right), \label{eq:catstate}
\ee
where we use the notation $\ket{\alpha, \xi} = \hat{D}(\alpha)\hat{S}(\xi)\ket{0}$ with the displacement operator $\hat{D}(\alpha) = e^{\alpha\create - \alpha^*\destroy}$. A projective measurement of the spin performed in the $\hat{\sigma}_z$ basis gives the probability of being $\ket{\downarrow}$ as $P(\downarrow) = (1 + X)/2$, where $X = \bra{\alpha, \xi}-\alpha, \xi\rangle = \bra{-\alpha, \xi} \alpha, \xi\rangle$ gives the overlap  between the two displaced motional states, which can be written as
\be
X(\alpha, \xi) = e^{-2 |\alpha|^2 \left(\exp(2 r)\cos^2(\Delta\phi)  + \exp(-2 r) \sin^2(\Delta\phi) \right) }  \label{eq:overlap}
\ee
where $\Delta\phi = \arg(\alpha) - \phi_s/2$. When $\Delta\phi = 0$, the SDF is aligned with the squeezed quadrature of the state, while for $\Delta \phi = \pi/2$, the SDF is aligned with the anti-squeezed quadrature. At displacements for which $X$ gives a measurable signal, monitoring the spin population as a function of the force duration $\tau$ for different choices of $\Delta\phi$ allows us to characterise the spatial variation of the initial squeezed wavepacket \cite{05Zeng, 10Gerritsma, 12Casanova}. For values of $|\alpha|^2$ which are greater than the wavepacket variance along the direction of the force, the state in equation (\ref{eq:catstate}) is a distinct superposition of squeezed wavepackets which have overlap close to zero and are entangled with the internal state. For $r = 0$ (no squeezing) the state reduces to the familiar ``Schr{\"o}dinger's cat'' states which have been produced in previous work \cite{96Monroe, 07McDonnell, 04Haljan, 13Wineland}. For $r>0$ the superposed oscillator states are the displaced-squeezed states \cite{81Caves, 76Yuen}.

\begin{figure}[b!]
\includegraphics[width = \columnwidth]{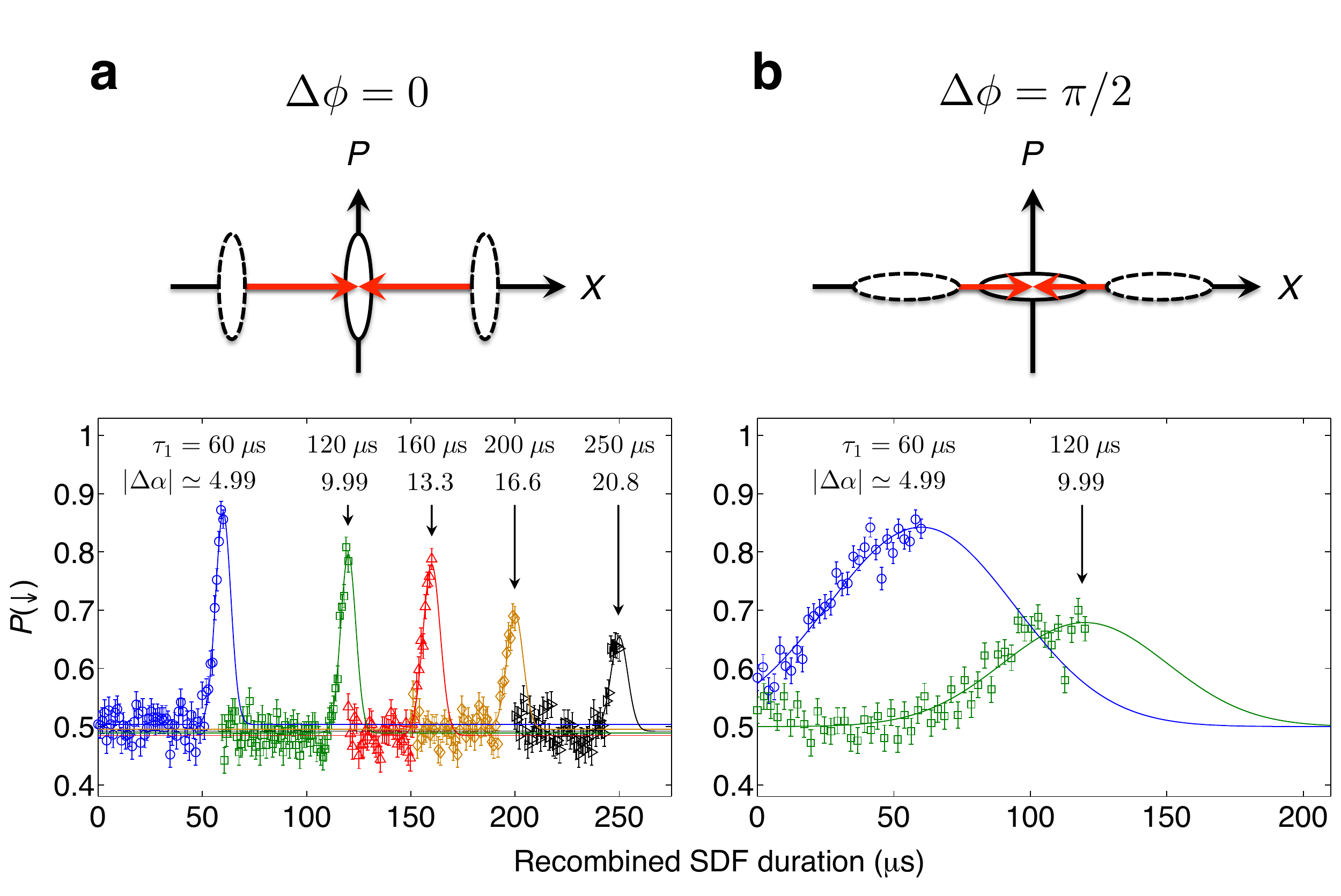}
\caption{\textbf{Revival of the spin coherence.} Spin populations as a function of the duration of the second SDF pulse with the spin phase shifted by $\pi$ relative to the first pulse. \textbf{a}, Forces parallel to the squeezed quadrature. \textbf{b}, Forces parallel to the anti-squeezed quadrature. In all cases an increase in the spin population is seen at the time when the two motional states are overlapped, which corresponds to the time $\tau_1$ used for the first SDF pulse. The value of $\tau_1$ and the corresponding $|\Delta\alpha|$ calculated from the measured Rabi frequency are written above the revival of each dataset. The fractional error on the mean of each of the estimated $|\Delta\alpha|$ is approximately 3\%. The solid lines are fitted curves using the same form as using in the fits in Fig. 1 with the overlap function $X(\delta \alpha, \xi)$. The obtained values of $r$ are consistent with the data in Fig. 1. The definition of error bars is the same as in Fig. 1.}
\label{fig:catreturn}
\end{figure}

The experiments use a single trapped $^{40}$Ca$^+$ ion, which mechanically oscillates on its axial vibrational mode with a frequency close to $\omega_z/(2 \pi)$ = 2.1~MHz. This mode is well resolved from all other modes. We encode a pseudo-spin system in the internal electronic states $\ket{\downarrow} \equiv \ket{S_{1/2}, M_J = 1/2}$ and $\ket{\uparrow} \equiv \ket{D_{5/2}, M_J = 3/2}$. All coherent manipulations, including the squeezed state preparation and the SDF, make use of the quadrupole transition between these levels at 729~nm, with a Lamb-Dicke parameter of $\eta \simeq 0.05$ for the axial mode. This is small enough that the experiments are well described using the Lamb-Dicke approximation (a discussion of this approximation is given in the Methods) \cite{98Wineland2}.

\begin{figure*}[ht!]
\includegraphics[width = \textwidth]{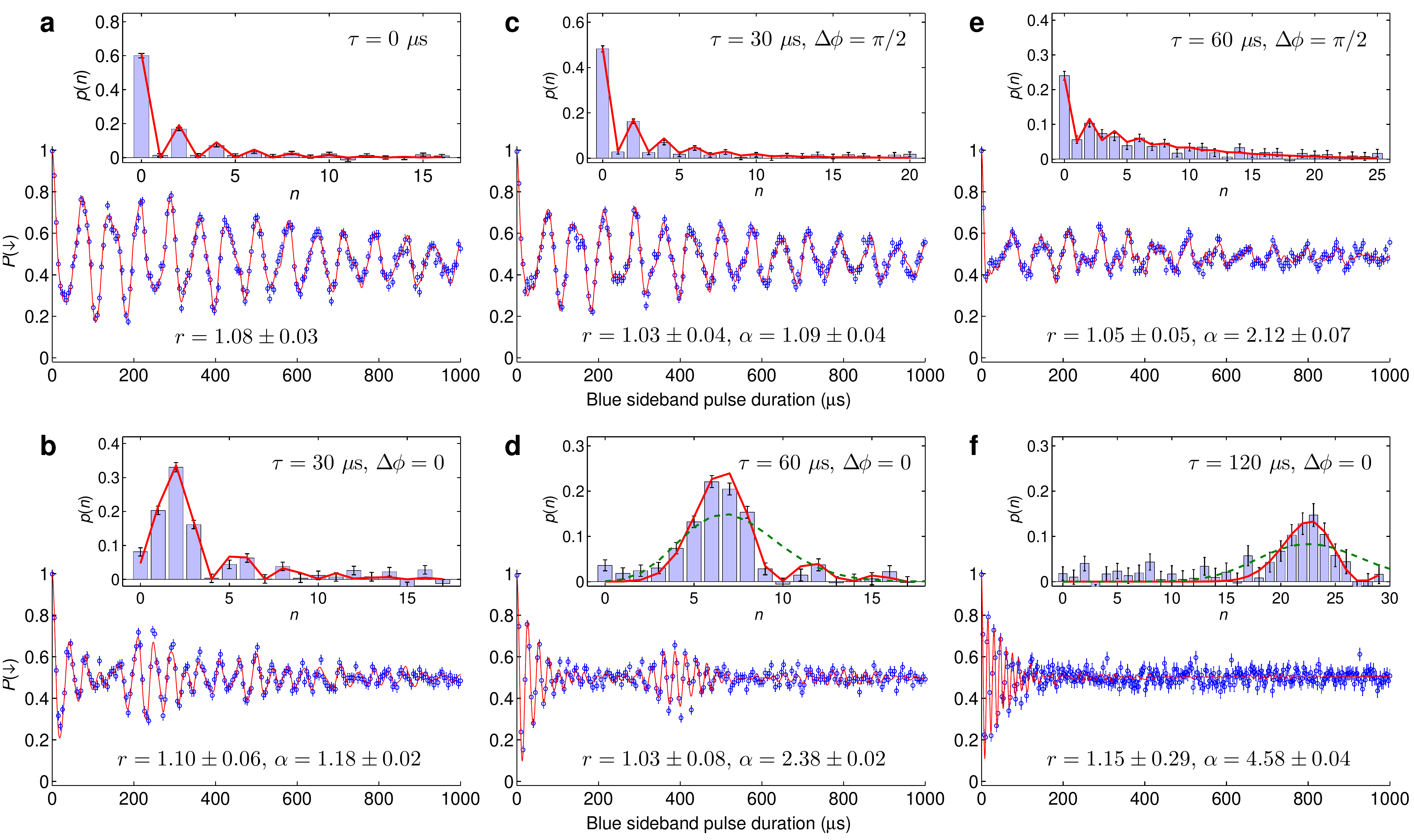}
\caption{\textbf{Evolution of displaced-squeezed state mixtures:} The observed blue-sideband oscillations and the corresponding number state probability distributions for the SDF applied along the two principal axes of the squeezed state and with different durations. \textbf{a}, Initial squeezed vacuum state. \textbf{b, d, f}, Forces parallel to the squeezed quadrature. \textbf{c, e}, Forces parallel to the anti-squeezed quadrature. For $\tau = 30~\mu$s the obtained parameters are consistent within statistical errors. For $\tau = 60~\mu$s the displacement along the anti-squeezed quadrature (\textbf{e}) results in a large spread in the number state probability distribution, with the result that in the fitting $r$ and $\alpha$ are positively correlated - the errors stated do not take account of this. We think that this accounts for the apparent discrepancy between the values of $r$ and $\alpha$ obtained for $\tau = 60~\mu$s. The green-dashed line in the inset of \textbf{d} and \textbf{f} is the Poisson distribution for the same $\langle n \rangle$ as the created displaced-squeezed state mixture, which is given by $\langle n \rangle = |\alpha|^2 + {\rm sinh}^2r$ \cite{81Caves}. The definition of error bars is the same as in Fig. 1.}
\label{fig:displacedsqueezed}
\end{figure*}

We apply the SDF directly after the squeezed vacuum state has been prepared by reservoir engineering and the internal state has been prepared in $\ket{\downarrow}$ by optical pumping (in the ideal case, the ion is already in the correct state and this step has no effect). Figure 1 shows the results of measuring $\langle \hat{\sigma}_z \rangle$ after applying displacements along the two principal axes of the squeezed state alongside the same measurement made using an ion prepared in the motional ground state. In order to extract relevant parameters regarding the SDF and the squeezing, we fit the data using $P(\downarrow) = (A + B X(\alpha, \xi))/2$, where the parameters $A$ and $B$ account for experimental imperfections such as shot-to-shot magnetic field fluctuations (Methods). Fitting the ground state data with $r$ fixed to zero allows us to extract $\Omega/(2 \pi) = 13.25 \pm 0.40$~kHz (here and in the rest of the paper, all errors are given as s.e.m.). We then fix this in performing independent fits to the squeezed-state data for $\Delta \phi = 0$ and $\Delta \phi = \pi/2$. Each of these fits allows us to extract an estimate for the squeezing parameter $r$. For both the squeezed and anti-squeezed quadratures we obtain consistent values with a mean of $r = 1.08 \pm 0.03$, corresponding to 9.4~dB reduction in the squeezed quadrature variance. The inset shows the spin population as a function of the SDF phase $\phi_D$ with the SDF duration fixed to 20~$\mu$s. This is also fitted using the same equation described above, and we obtain $r = 1.13\pm0.03$.

The loss of overlap between the two wavepackets indicates that a SWES has been created. In order to verify that these states are coherent superpositions, we recombine the wavepackets by applying a second ``return'' SDF pulse for which the phase of both the red and blue sideband laser frequency components is shifted by $\pi$ relative to the first. This reverses the direction of the force applied to the motional states for both the $\ket{+}$ and $\ket{-}$ spin states. In the ideal case a state displaced to $\alpha(\tau_1)$ by a first SDF pulse of duration $\tau_1$ has a final displacement of $\delta\alpha = \alpha(\tau_1) - \alpha(\tau_2)$ after the return pulse of duration $\tau_2$. For $\tau_1 = \tau_2$, $\delta\alpha = 0$ and the measured probability of finding the spin state in $\ket{\downarrow}$ is 1. In the presence of decoherence and imperfect control, the probability with which the ion returns to the $\ket{\downarrow}$ state will be reduced. In Fig. 2 we show revivals in the spin coherence for the same initial squeezed vacuum state as was used for the data in Fig. 1. The data include a range of different $\tau_1$. For the data where the force was applied along the squeezed axis of the state ($\Delta \phi = 0$), partial revival of the coherence is observed for SDF durations up to $250~\mu$s. For $\tau_1 = 250~\mu$s the maximum separation of the two distinct oscillator wavepackets is $|\Delta\alpha| > 19$, which is 56 times the r.m.s width of the squeezed wavepacket in phase space. The amplitude of revival of this state is similar to what we observe when applying the SDF to a ground state cooled ion. The loss of coherence as a function of the displacement duration is consistent with the effects of magnetic-field induced spin dephasing and motional heating \cite{00Turchette2,13Hempel}. When the force is applied along the anti-squeezed quadrature ($\Delta \phi = \pi/2$), we observe that the strength of the revival decays more rapidly than for displacements with $\Delta\phi = 0$. Simulations of the dynamics using a quantum Monte-Carlo wavefunction approach including sampling over a magnetic field distribution indicate that this is caused by shot-to-shot fluctuations of the magnetic field (Methods).

\begin{figure}[t!]
\includegraphics[width = 0.95\columnwidth]{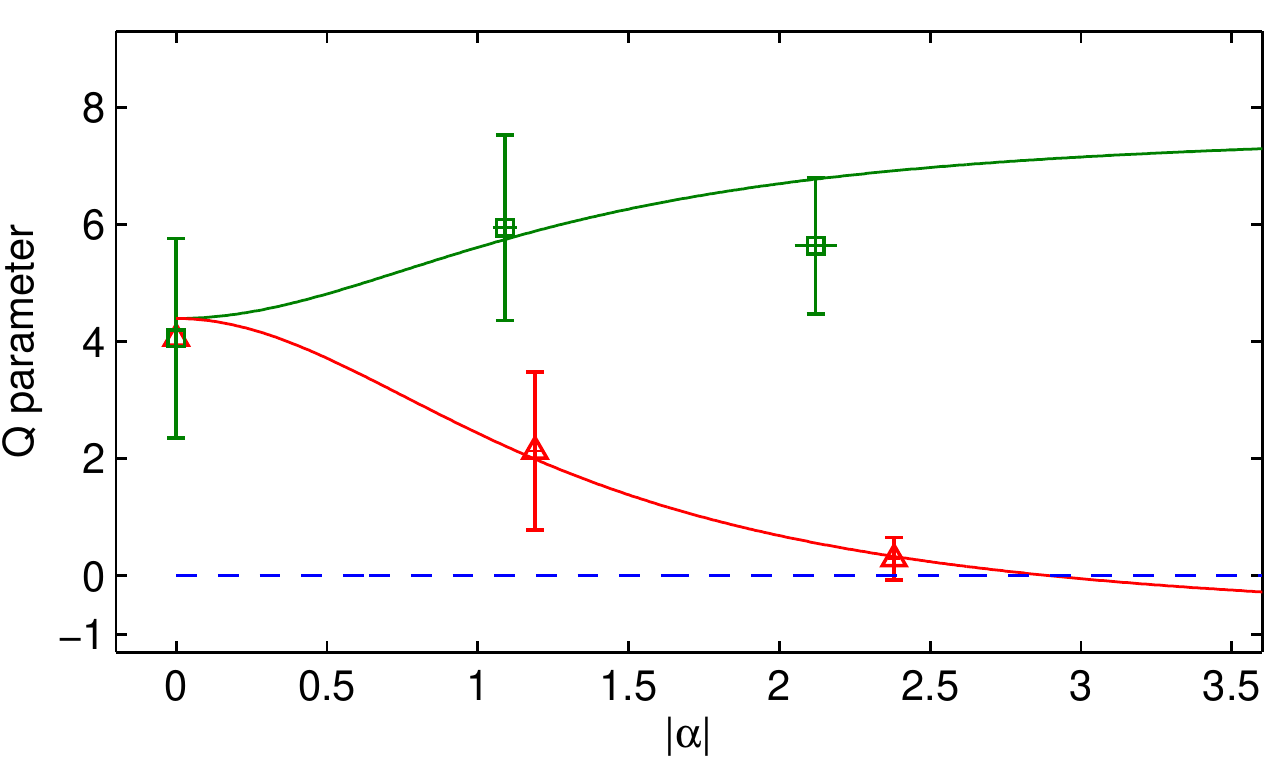}
\caption{\textbf{Mandel $Q$ parameter for the displaced-squeezed states:} Shown are the results for displacements along the squeezed quadrature (red triangles) and the anti-squeezed quadrature (green squares). All the values are calculated from the experimental data given in Fig. 3, taking the propagation of error into account. The solid lines are theoretical curves for displacements along the squeezed (red) and anti-squeezed (green) quadratures of an initial state with $r = 1.08$. The values of $|\alpha|$ are obtained from fits to the respective $p(n)$ (Fig. 3), with error bars comparable to the size of the symbol. The point at $|\alpha| = 0$ is the squeezed vacuum state.}
\label{fig:variance}
\end{figure}

We are also able to monitor the number state distributions of the motional wavepackets as a function of the duration of the SDF. This provides a second measurement of the parameters of the SDF and the initial squeezed wavepacket which has similarities with the homodyne measurement used in optics \cite{97Breitenbach,07Ourjoumtsev}. In order to do this, we optically pump the spin state into $\ket{\downarrow}$ after applying the SDF. This procedure destroys the phase relationship between the two motional wavepackets, resulting in the mixed oscillator state $\hat{\rho}_{\rm mixed} = \left(\ket{\alpha, \xi }\bra{\alpha, \xi} + \ket{-\alpha, \xi}\bra{-\alpha, \xi}\right)/2$ (we estimate the photon recoil during optical pumping results in a reduction in the fidelity of our experimental state relative to $\hat{\rho}_{\rm mixed}$ by $< 3\%$, which would not be observable in our measurements). The two parts of this mixture have the same number state distribution, which is that of a displaced-squeezed state \cite{81Caves, 76Yuen}. In order to extract this distribution, we drive Rabi oscillations on the blue-sideband transition \cite{96Meekhof} and monitor the subsequent spin population in the $\hat{\sigma}_z$ basis. Figure 3 shows this evolution for SDF durations of $\tau = 0$, 30, 60 and 120~$\mu$s. For $\tau = 30$ and $60~\mu$s, the results from displacements applied parallel to the two principal axes of the squeezed state are shown ($\Delta\phi = 0$ and $\pi/2$). We obtain the number state probability distribution $p(n)$ from the spin state population by fitting the data using a form $P(\downarrow) = b t  + \frac{1}{2}\sum_n p(n) (1 +  e^{-\gamma t}\cos(\Omega_{n, n + 1} t))$, where $t$ is the blue-sideband pulse duration, $\Omega_{n,n+1}$ is the Rabi frequency for the transition between the $\ket{\downarrow}\ket{n}$ and $\ket{\uparrow}\ket{n+1}$ states and $\gamma$ is a phenomenological decay parameter \cite{96Meekhof, 03Leibfried2}. The parameter $b$ accounts for gradual pumping of population into the state $\ket{\uparrow}\ket{0}$  due to frequency noise on our laser \cite{00Fidio,14Kienzler}. It is negligible when $p(0)$ is small. The resulting $p(n)$ are then fitted using the theoretical form for the displaced-squeezed states (Methods). The number state distributions show a clear dependence on the phase of the force, which is also reflected in the spin population evolution. Figure 4 shows the Mandel $Q$ parameters of the experimentally obtained number state distributions, defined as $Q = \langle (\Delta n)^2\rangle/\langle n\rangle - 1$ in which $\langle (\Delta n)^2\rangle$ and $\langle n\rangle$ are the variance and mean of $p(n)$ respectively \cite{79Mandel}. The solid lines are the theoretical curves given by Caves \cite{81Caves} for $r = 1.08$, and are in agreement with our experimental results. For displacements along the short axis of the squeezed state (Fig. 3), the collapse and revival behaviour of the time evolution of $P(\downarrow)$ is reminiscent of the Jaynes-Cummings Hamiltonian applied to a coherent state \cite{BkHaroche}, but it exhibits a higher number of oscillations before the ``collapse'' for a state of the same $\langle n\rangle$. This is surprising since the statistics of the state is not sub-Poissonian. We attribute this to the fact that this distribution is more peaked than that of a coherent state with the same  $\langle n \rangle$, which is obvious when the two distributions are plotted over one another (Figs. 3(d) and 3(f)). The increased variance of the squeezed state then arises from the extra populations at high $n$, which are too small to make a visible contribution to the Rabi oscillations. For the squeezing parameter in our experiments sub-Poissonian statistics would only be observed for $|\alpha|>3$. For $\tau = 120~\mu$s we obtain a consistent value of $r$ and $|\alpha| = 4.6$ only in the case where we include a fit parameter for scaling of the theoretical probability distribution, obtaining a fitted scaling of $0.81\pm0.10$ (Methods). The reconstruction of the number state distribution is incomplete, since we cannot extract populations with $n>29$ due to frequency crowding in the $\sqrt{n+1}$ dependence of the Jaynes-Cummings dynamics. As a result, we do not include these results in Fig. 4. Measurement techniques made in a squeezed-state basis \cite{14Kienzler} could avoid this problem, however these are beyond our current experimental capabilities for states of this size.

We have generated entangled superposition states between the internal and motional states of a single trapped ion in which the superposed motional wavepackets are of a squeezed Gaussian form. These states present new possibilities both for metrology and for continuous variable quantum information. In an interferometer based on SC states separated by $|\Delta \alpha|$, the interference contrast depends on the final overlap of the re-combined wavepackets. Fluctuations in the frequency of the oscillator result in a reduced overlap, but this effect can be improved by a factor $\exp{\left[-\left|\Delta \alpha \right|^2(e^{-2r} -1)/2\right]}$ if the wavepackets are squeezed in the same direction as the state separation (Methods). In quantum information with continuous variables, the computational basis states are distinguishable because they are separated in phase space by $|\Delta\alpha|$ and thus do not overlap \cite{12Weedbrook, 00Gottesman2, 02Bartlett}. The decoherence times of such superpositions typically scale as $1/|\Delta \alpha|^2$ \cite{00Turchette2}. The use of states squeezed along the displacement direction reduces the required displacement for a given overlap by $e^{r}$, increasing the resulting coherence time by $e^{2 r}$ which is a factor of 9 in our experiments. We therefore expect these states to open up new possibilities for quantum state engineering and control.\\

\textbf{Acknowledgement:} We thank Joseba Alonso and Florian Leupold for comments on the manuscript and thank Florian Leupold, Frieder Lindenfelser, Joseba Alonso, Martin Sepiol, Karin Fisher, and Christa Fl{\"u}hmann for contributions to the experimental apparatus. We acknowledge support from the Swiss National Science Foundation under grant number $200021\_134776$, and through the National Centre of Competence in Research for Quantum Science and Technology (QSIT).\\

\textbf{Author Contributions:} Experimental data were taken by H.Y.L., D.K., and L.d.C., using an apparatus primarily built up by D.K., H.Y.L., and B.C.K., and with significant contributions from L.d.C., V.N., and M.M.. Data analysis was performed by H.Y.L. and J.P.H.. The paper was written by J.P.H. and H.Y.L., with input from all authors. The work was conceived by J.P.H..\\

The authors have no competing financial interests.

\section*{Methods}

\begin{figure*}[ht!]
\includegraphics[width = 0.9\textwidth]{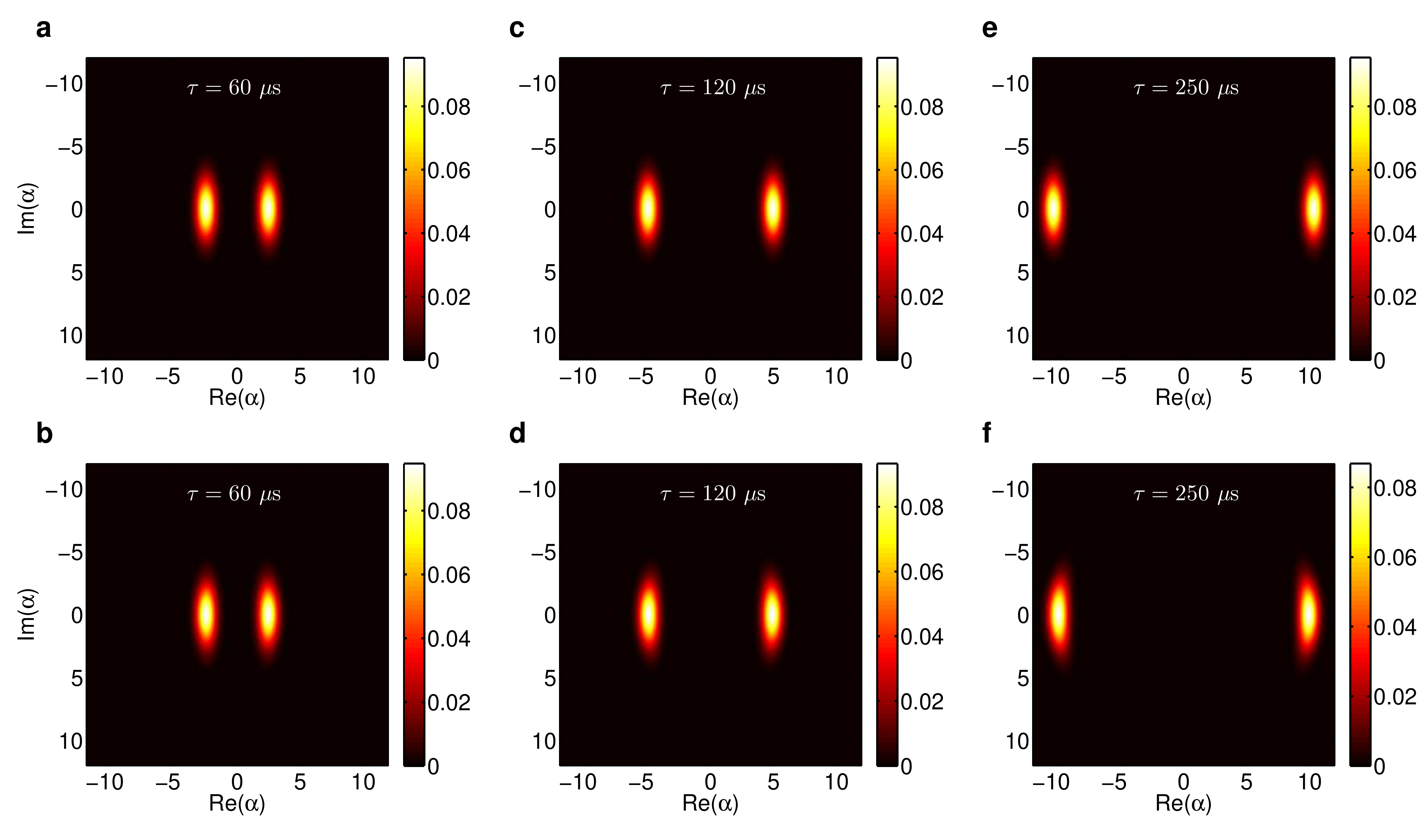}
\caption{\textbf{Quasi-probability distributions for displaced-squeezed states in phase space using LDA and non-LDA:} \textbf{a, c, e}, The simulation results using LDA with different SDF durations. \textbf{b, d, f}, The results simulated using the full Hamiltonian.
}
\label{fig:squeezed_short_comparison}
\end{figure*}

\textbf{Experimental details.} The experiments make use of a segmented linear Paul trap with an ion-electrode distance of $\approx 185~\mu$m. Motional heating rates from the ground state for a calcium ion in this trap have been measured to be 10~$\pm$~1~quanta\ s$^{-1}$, and the coherence time for the number state superposition $(\ket{0}+ \ket{1})/\sqrt{2}$ has been measured to be 32 $\pm$ 3~ms.

The first step of each experimental run involves cooling all modes of motion of the ion close to the Doppler limit using laser light at 397 and 866~nm. The laser beam used for coherent control of the two-level pseudo-spin system addresses the narrow-linewidth transition $\ket{\downarrow}\equiv\ket{S_{1/2}, M_J = 1/2} \leftrightarrow \ket{\uparrow}\equiv\ket{D_{5/2}, M_J = 3/2}$ at 729~nm. This transition is resolved by 200~MHz from all other internal state transitions in the applied magnetic field of 119.6~Gauss (G). The State-Dependent Forces (SDFs) and the reservoir engineering \cite{14Kienzler} in our experiment require the application of a bichromatic light field. We generate both frequency components using Acousto-Optic-Modulators (AOMs) starting from a single laser stabilized to an ultra-high-finesse optical cavity with a resulting linewidth $< 600$~Hz (at which point magnetic field fluctuations limit the qubit coherence). We apply pulses of 729~nm laser light using a double-pass AOM to which we apply a single radio-frequency tone, followed by a single-pass AOM to which two radio-frequency tones are applied. Following this second AOM, both frequency components are coupled into the same single-mode fibre before delivery to the ion. The double-pass AOM is used to switch on and off the light. Optical pumping to $\ket{\downarrow}$ is implemented using a combination of linearly polarized light fields at 854~nm, 397~nm and 866~nm. The internal state of the ion is read out by state-dependent fluorescence using laser fields at 397~nm and 866~nm.

The 729~nm laser beam enters the trap at 45 degrees to the $z$ axis of the trap resulting in a Lamb-Dicke parameter of $\eta \simeq 0.05$ for the axial mode. For this Lamb-Dicke parameter, we have verified whether for displacements up to $|\alpha| = 9.75$ the dynamics can be well described with the Lamb-Dicke Approximation (LDA). We simulate the wavepacket dynamics using the interaction Hamiltonian with and without LDA. In the simulation, we apply the SDF to an ion prepared in $\ket{\downarrow}\ket{\xi}$. The interaction Hamiltonian for a single trapped ion coupled to a single-frequency laser field can be written as \cite{03Leibfried2}
\begin{eqnarray*}
\hat{H}_{I} = \frac{\hbar}{2}\Omega_0\hat{\sigma}_+ {\rm exp}{\{i\eta(\destroy e^{-i\omega_z t} + \create e^{i\omega_z t})\} e^{i(\phi -\delta t)}} + {\rm h.c.},
\end{eqnarray*}
where $\Omega_0$ is the interaction strength, $\hat{\sigma}_+ = \ket{\uparrow}\bra{\downarrow}$, $\destroy$ and $\create$ are motional annihilation and creation operators, $\omega_z$ is the vibrational frequency of the ion, $\phi$ is the phase of the laser, and $\delta = \omega_l - \omega_a$ the detuning of the laser from the atomic transition. In the laboratory, the application of the SDF involves simultaneously driving both the blue and red sideband transitions resonantly resulting in the Hamiltonian $\hat{H}_{\rm tot} = \hat{H}_{\rm bsb} + \hat{H}_{\rm rsb}$, where $\delta = \omega_z$ in $\hat{H}_{\rm bsb}$ and $\delta = -\omega_z$ in $\hat{H}_{\rm rsb}$. Starting from $\ket{\downarrow}\ket{\xi}$, the evolution of the state can not be solved analytically. We perform a numerical simulation in which we retain only the resonant terms in the Hamiltonian. Figure \ref{fig:squeezed_short_comparison} shows the quasi-probability distributions in phase space for chosen values of the SDF duration $\tau$. These are compared to results obtained using the LDA. For $\tau = 60~\mu$s both cases are similar, resulting in $|\alpha| \simeq 2.4$. For $\tau = 250~\mu$s the squeezed state wavepackets are slightly distorted and the displacement is 4\% smaller for the full simulation than for the LDA form. Considering the levels of error arising from imperfect control and decoherence for forces of this duration, we do not consider this effect to be significant in our experiments.

\begin{figure}[b]
\includegraphics[width = 0.97\columnwidth]{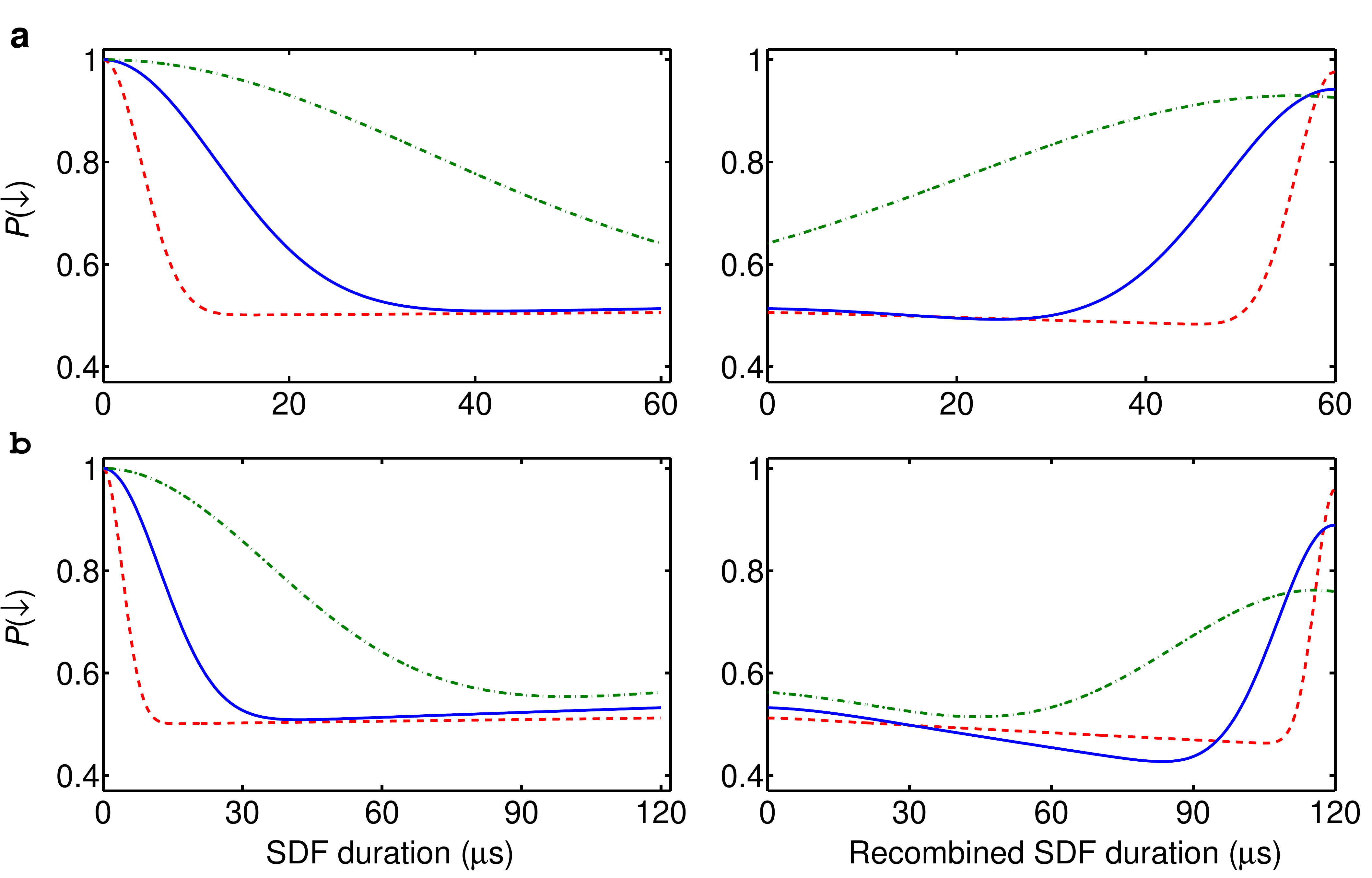}
\caption{\textbf{Coherence of cat states with fixed magnetic field noise:} The magnetic-field-induced energy-level-shift of 1.5~kHz is used in this simulation. \textbf{a}, The duration of both SDF pulses is 60~$\mu$s. \textbf{b}, The duration of both SDF pulses is 120~$\mu$s. Red-dashed and green-dash-dot curves show the SDF aligned along the squeezed and anti-squeezed quadratures. The blue trace is for the SDF applied to a ground state cooled ion.
}
\label{fig:fixed_noise}
\end{figure}

\begin{figure}[t]
\includegraphics[width = 0.97\columnwidth]{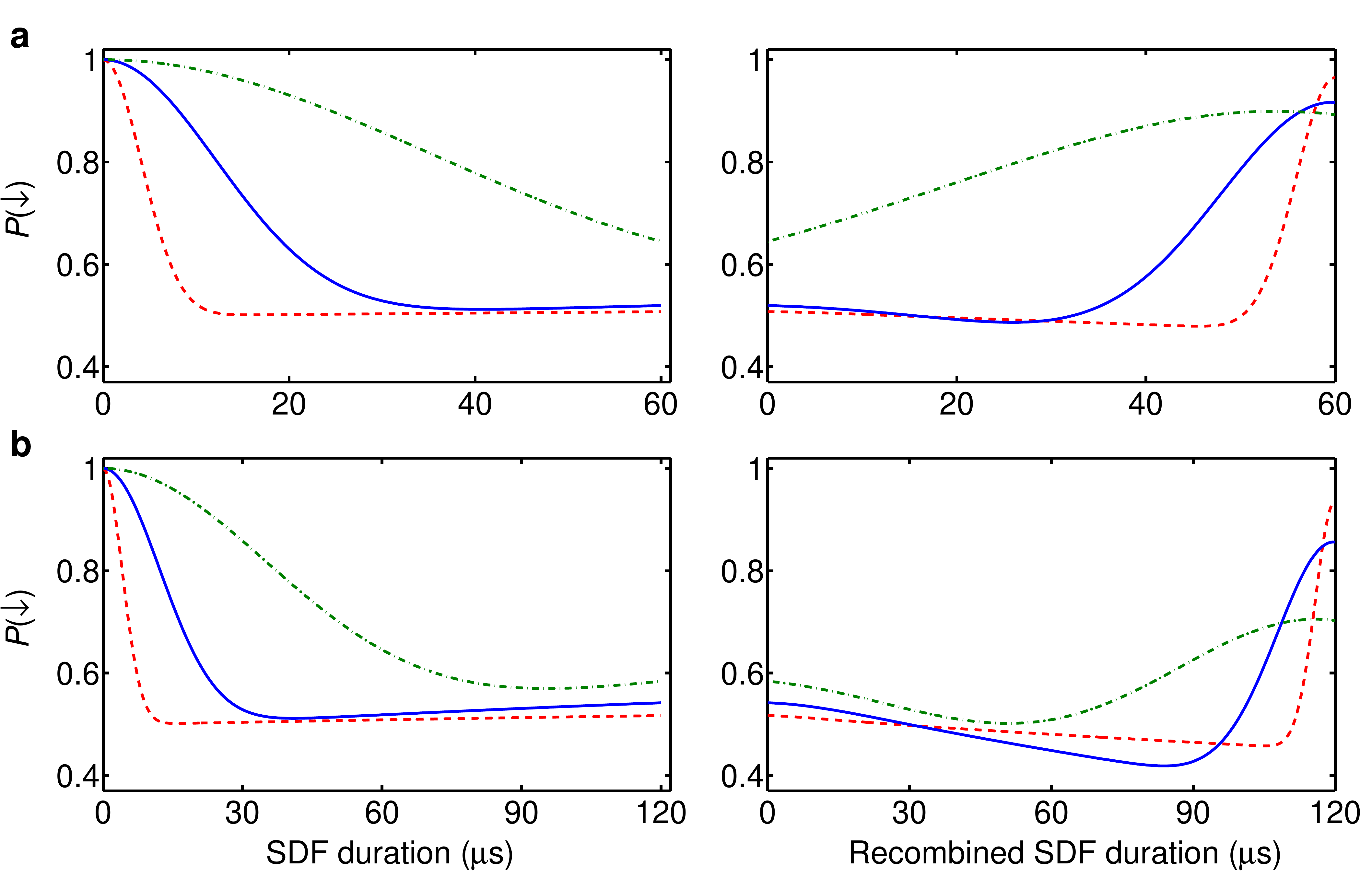}
\caption{\textbf{Coherence of cat states with a magnetic field fluctuation distribution:} Assuming the magnetic field exhibits a 50-Hz sinusoidal pattern with an amplitude of 2.2~mG, this plot shows the simulation results by taking an average over 100 samples on the field distribution. \textbf{a}, The duration of both SDF pulses is 60~$\mu$s. \textbf{b}, The duration of both SDF pulses is 120~$\mu$s. Definitions of the curve specification are the same as Extended Data Fig. 2.
}
\label{fig:sampling_noise}
\end{figure}

\textbf{Simulations for the coherence of SWESs.} After creating SWESs, we deduce that coherence is retained throughout the creation of the state by applying a second SDF pulse to the ion, which recombines the two separated wavepackets and disentangles the spin from the motion. The revival in the spin coherence is not perfect due to decoherence and imperfect control in the experiment. One dominant source causing decoherence of the superpositions is spin decoherence due to magnetic field fluctuations. We have performed quantum Monte-Carlo wavefunction simulations to investigate the coherence of the SWES in the presence of such a decoherence mechanism. We simulate the effect of a sinusoidal fluctuation of the magnetic field on a timescale long compared to the duration of the coherent control sequence, which is consistent with the noise which we observe on our magnetic field coil supply (at 10 and 110~Hz) and from ambient fluctuations due to electronics equipment in the room. The amplitude of these fluctuations is set to $\simeq$ 2.2~mG, giving rise to the spin coherence time of 180~$\mu$s which we have measured using Ramsey experiments on the spin alone. Since the frequency of fluctuations is slow compared to the sequence length, we fix the field for each run of the simulation, but sample its value from a probability distribution derived from a sinusoidal oscillation. In Fig. \ref{fig:fixed_noise} we show the effect of a single shot taken at a fixed qubit-oscillator detuning of 1.5~kHz, while in Fig. \ref{fig:sampling_noise} we show the average over the distribution. In both figures results are shown for the SDF applied along the two principal axes of the squeezed vacuum state as well as for the motional ground state using force durations of 60 and 120~$\mu$s. We also show the results of applying the second SDF pulse resulting in partial revival of the spin coherence. It can be clearly seen that when the SDF is applied along the anti-squeezed quadrature, the strength of the revival decays more rapidly, and $P(\downarrow)$ oscillates around 0.5. This effect can be seen in the data shown in Fig. 2 of the main article.

\textbf{Number state probability distributions for the displaced-squeezed state.} For Fig. 3 in the main article, we characterise the probability distribution for the number states of the oscillator. This is performed by driving the blue-sideband transition $\ket{\downarrow}\ket{n}\leftrightarrow \ket{\uparrow}\ket{n + 1}$ and fitting the obtained spin population evolution using
\be
P(\downarrow) = bt + \frac{1}{2}\sum_n p(n) (1 +  e^{-\gamma t}\cos(\Omega_{n, n + 1} t)) \label{eq:population} ,
\ee
where $t$ is the blue-sideband pulse duration, $p(n)$ are the number state probabilities for the motional state we concern, and $\gamma$ is an empirical decay parameter \cite{96Meekhof, 03Leibfried2}. In the results presented here we do not scale this decay parameter with $n$ as was done by \cite{96Meekhof}. We have also fitted the data including such a scaling and see consistent results. The Rabi frequency coupling $\ket{\downarrow}\ket{n}$ to $\ket{\uparrow}\ket{n+1}$ is $\Omega_{n,n +1} = \Omega_0 |\bra{n} e^{i \eta (\create + \destroy)} \ket{n + 1}| = \Omega_0 e^{-\eta^2/2} \eta L^1_n(\eta^2)/\sqrt{n + 1}$. For small $n$, this scales as $\sqrt{n+1}$, but since the states include significant populations at higher $n$ we use the complete form including the generalized Laguerre polynomial $L_n^1(x)$. The parameter $b$ in the first term accounts for a gradual pumping of population into the state $\ket{\uparrow}\ket{0}$ which is not involved in the dynamics of the blue-sideband pulse \cite{00Fidio,14Kienzler}. This effect is negligible when $p(0)$ is small.

After extracting $p(n)$ from $P(\downarrow)$, we fit it using the number state probability distribution for the displaced-squeezed state \cite{bkKnight},
\begin{widetext}
\begin{eqnarray*}
p(n) = \kappa {\:} \frac{(\frac{1}{2} {\rm tanh}{\,}r )^n}{n!{\:}{\rm cosh}{\,}r } {\rm exp}\left[  -|\alpha|^2 - \frac{1}{2}(\alpha^{*2}e^{i\phi_s} + \alpha^2 e^{-i\phi_s}){\:}{\rm tanh}{\:}r \right] \left| H_n\left[  \frac{\alpha{\:}{\rm cosh}{\,}r + \alpha^*e^{i \phi_s}{\rm sinh}{\,}r}{\sqrt{e^{i\phi_s}{\:}{\rm sinh}{\,}2r}} \right]\right|^2,\\
\end{eqnarray*}
\end{widetext}
where $\kappa$ is a constant which accounts for the infidelity of the state during the application of SDF and the $H_n(x)$ are the Hermite polynomials. The direction of the SDF is aligned along either the squeezing quadrature or the anti-squeezing quadrature of the state. Therefore, we set ${\rm arg}(\alpha) = 0$ and fix $\phi_s = 0$ and $\pi$ for fitting the data of the short axis and the long axis of the squeezed state, respectively. This allows us to obtain the values of $r$ and $|\alpha|$ for the state we created. For the cases of smaller displacements (from Figs. 3(a) to (e) in the main article), we set $\kappa = 1$. For the data set of $|\alpha| \simeq 4.6$ (Fig. 3(f) in the main article), $\kappa$ is a fitting parameter which gives us a value of $0.81 \pm 0.1$. We note that in this case $4\%$ of the expected population lies above $n = 29$ but we are not able to extract these populations from our data.

The Mandel $Q$ parameter \cite{79Mandel}, defined as
\begin{eqnarray*}
Q = \frac{\langle(\Delta n)^2\rangle - \langle n\rangle}{\langle n\rangle}.
\end{eqnarray*}
where $\langle n\rangle$ and $\langle(\Delta n)^2\rangle$ are the mean and variance of the probability distribution. For a displaced-squeezed state these are given by Caves \cite{81Caves} as
\begin{eqnarray*}
\langle(\Delta n)^2\rangle &=& \left| \alpha{\:}{\rm cosh}{\,}r - \alpha^* e^{i\phi_s} {\rm sinh}{\:}r \right|^2 + 2{\:}{\rm cosh}^2r{\:}{\rm sinh}^2r, \\
\langle n\rangle &=& |\alpha|^2 + {\rm sinh}^2r.\\
\end{eqnarray*}
These forms were used to produce the curves given in Fig. 4 of the main article.

\begin{figure}[ht!]
\includegraphics[width = 0.95\columnwidth]{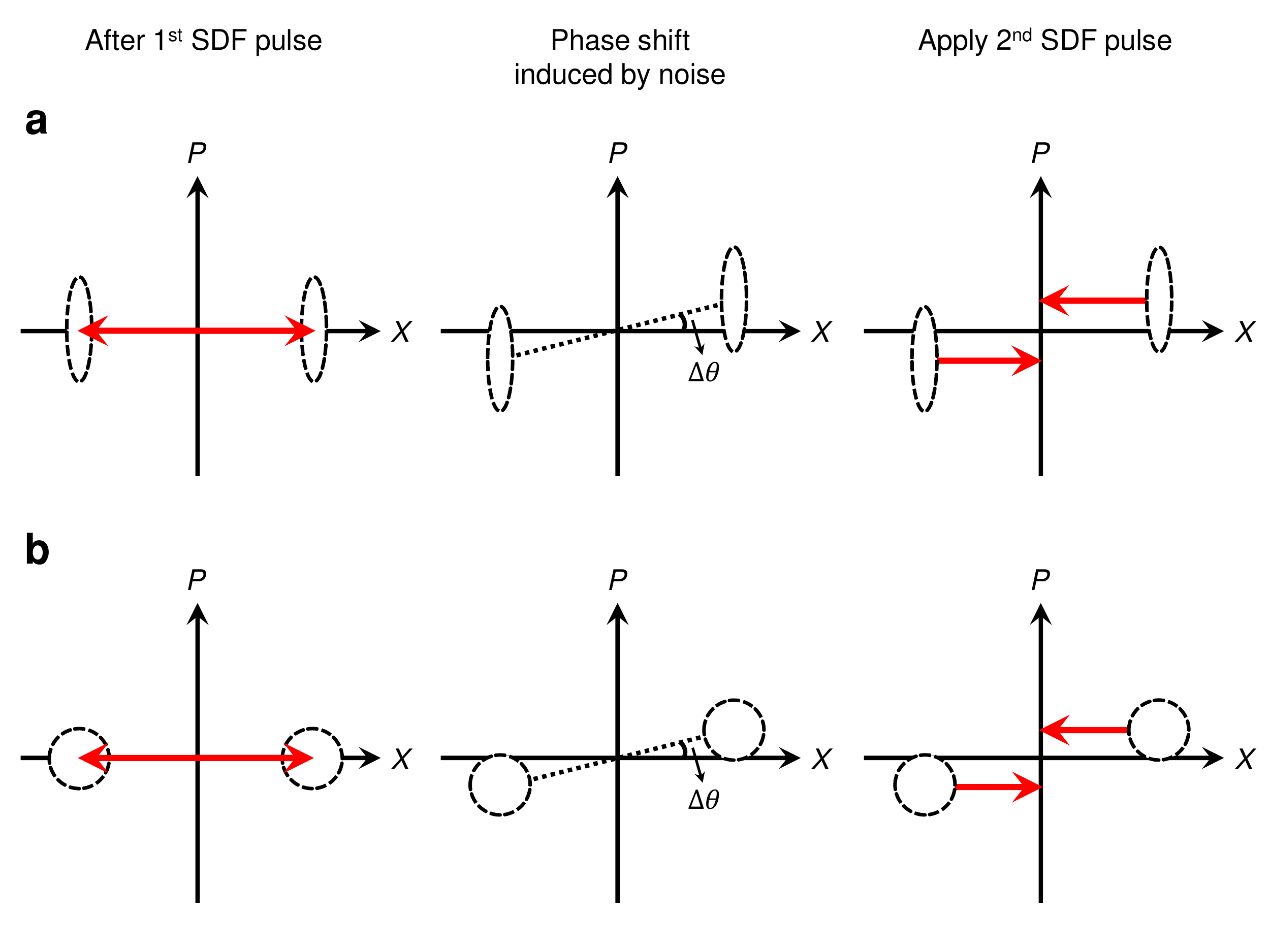}
\caption{\textbf{Possible application of using SWESs for interferometry:} \textbf{a}, Use of squeezed state wavepackets. \textbf{b}, Use of ground state wavepackets. The first SDF pulse is used to create a spin-motion entangled state. In the middle, a small phase shift $\Delta\theta$ is induced by shot-to-shot fluctuation in the oscillator frequency before the application of the second SDF pulse, which recombines the two distinct oscillator wavepackets.
}
\label{fig:intermerometry}
\end{figure}

\textbf{Applications of SWESs.} The SWES may offer new possibilities for sensitive measurements which are robust against certain types of noise. An example is illustrated in Fig. \ref{fig:intermerometry} where we compare an interferometry experiment involving the use of a SWES versus a more standard Schr\"odinger's cat state based on coherent states. In both cases the superposed states have a separation of $|2\alpha|$ obtained using a SDF. For the SWES this force is aligned along the squeezed quadrature of the state. The interferometer is closed by inverting the initial SDF, resulting in a residual displacement which in the ideal case is zero. One form of noise involves a shot-to-shot fluctuations in the oscillator frequency. On each run of the experiment, this would result in a small phase shift $\Delta\theta$ arising between the two superposed motional states. As a result, after the application of the second SDF pulse the residual displacement would be $\alpha_R = 2 i \alpha \sin(\Delta \theta/2)$, which corresponds to the states being separated along the $P$ axis in the rotating-frame phase space. The final state of the system would then be $\ket{\psi(\alpha_R)}$ with a corresponding state overlap given by $X(\alpha_R,\xi)$. Therefore the contrast will be higher for the SWES (Extended Data Fig. 4(a)) than for the coherent Schr\"odinger's cat state (Extended Data Fig. 4(b)) by a factor
\begin{eqnarray*}
\exp{\left[-2\left|\alpha_R\right|^2(e^{-2r} -1)\right]}.
\end{eqnarray*}
While in our experiments other sources of noise dominate, in other systems such oscillator dephasing may be more significant.

\bibliography{./myrefs}

\end{document}